\documentclass[aip,jap
,reprint
]{revtex4-1}
\usepackage{graphicx}% Include figure files
\usepackage{txfonts}%[varg]

\usepackage{mathrsfs}
\usepackage{amsmath}
\usepackage{bm}

\renewcommand{\vec}[1]{\mathbf{#1}}      %vector --- boldsymbol
\allowdisplaybreaks
\begin{document}

% Use the \preprint command to place your local institutional report number 
% on the title page in preprint mode.
% Multiple \preprint commands are allowed.
%\preprint{}

\title{Graphene-based nanodynamometer} %Title of paper

% repeat the \author .. \affiliation  etc. as needed
% \email, \thanks, \homepage, \altaffiliation all apply to the current author.
% Explanatory text should go in the []'s, 
% actual e-mail address or url should go in the {}'s for \email and \homepage.
% Please use the appropriate macro for the type of information

% \affiliation command applies to all authors since the last \affiliation command. 
% The \affiliation command should follow the other information.

\author{N.A. Poklonski}
\email{poklonski@bsu.by}
\author{A.I. Siahlo}
\author{S.A. Vyrko}
\affiliation{Physics Department, Belarusian State University, pr.~Nezavisimosti 4, Minsk 220030, Belarus}

\author{A.M. Popov}
\affiliation{Institute of Spectroscopy, Fizicheskaya Str. 5, Troitsk, Moscow Region 142190, Russia}
\author{Yu.E. Lozovik}
\affiliation{Institute of Spectroscopy, Fizicheskaya Str. 5, Troitsk, Moscow Region 142190, Russia}
\affiliation{Moscow Institute of Physics and Technology, Institutskii pereulok 9, Dolgoprudny, Moscow Region 141701, Russia}

\author{I.V.~Lebedeva}
\affiliation{Moscow Institute of Physics and Technology, Institutskii pereulok 9, Dolgoprudny, Moscow Region 141701, Russia}
\affiliation{RRC ``Kurchatov Institute'', Kurchatov Sq. 1, Moscow 123182, Russia}
\affiliation{Kintech Lab Ltd, Kurchatov Sq. 1, Moscow 123182, Russia}

\author{A.A.~Knizhnik}
\affiliation{Kintech Lab Ltd, Kurchatov Sq. 1, Moscow 123182, Russia}
\affiliation{RRC ``Kurchatov Institute'', Kurchatov Sq. 1, Moscow 123182, Russia}

% Collaboration name, if desired (requires use of superscriptaddress option in \documentclass). 
% \noaffiliation is required (may also be used with the \author command).
%\collaboration{}
%\noaffiliation

\date{\today}

\begin{abstract}
A new concept of an electromechanical nanodynamometer based on the relative displacement of layers of bilayer graphene is proposed. In this nanodynamometer, force acting on one of the graphene layers causes the relative displacement of this layer and related change of conductance between the layers. Such a force can be determined by measurements of the tunneling conductance between the layers. Dependences of the interlayer interaction energy and the conductance between the graphene layers on their relative position are calculated within the first-principles approach corrected for van der Waals interactions and the Bardeen method, respectively. The characteristics of the nanodynamometer are determined and its possible applications are discussed.
\end{abstract}

%\pacs{72.80.Vp, 73.40.Gk, 81.07.Oj}% insert suggested PACS numbers in braces on next line
%\keywords{Bilayer graphene, Tunneling conductance, Nanoelectromechanical systems}%Use showkeys class option if keyword display desired

\maketitle %\maketitle must follow title, authors, abstract and \pacs

% Body of paper goes here. Use proper sectioning commands. 
% References should be done using the \cite, \ref, and \label commands
\section{Introduction}
Due to the unique electrical and mechanical properties, carbon nanostructures (fullerenes, carbon nanotubes and graphene) are considered as promising materials for the use in nanoelectromechanical systems (NEMS). Since the conductance of carbon nanotubes depends on the relative displacement of nanotube walls at the sub-nanometer scale,\cite{Grace04, Tunney06, PoklonskiHieu08} a set of nanosensors based on such a displacement was proposed.
This set includes a variable nanoresistor,\cite{LozovikMinogin03, YanZhou06} a strain nanosensor \cite{Bichoutskaia06} and a nanothermometer \cite{Bichoutskaia07, PopovBichoutskaia07} (see Ref.~\onlinecite{LozovikPopov07} for a review). 
A number of nanotube-based NEMS, such as a nanoresonator based on a suspended nanotube \cite{Sazonova04, Peng06} and a nanoaccelerometer based on a telescoping nanotube,\cite{Wang08, Kang09} were suggested as means for measurements of small forces and accelerations by detection of changes in the system capacitance. In addition to zero-dimensional and one-dimensional carbon nanostructures, fullerenes and carbon nanotubes, a novel two-dimensional carbon nanostructure, graphene, was discovered recently.\cite{Novoselov04} By analogy with NEMS based on carbon nanotubes, nanodevices based on graphene were proposed.\cite{Zheng08, Lebedeva11}
A nanoresonator based on flexural vibrations of suspended graphene was implemented.\cite{Bunch07} Similar to devices based on the dependence of conductance of carbon nanotubes on the relative displacement of nanotube walls, NEMS based on the dependence of conductance of graphene on the relative displacement of graphene layers can be considered.

In this paper, we propose a new concept of an electromechanical nanodynamometer based on the relative displacement of layers of bilayer graphene and investigate the operating characteristics of this sensor. The conceptual design of the nanodynamometer is shown in Fig.~1. The operation of the nanodynamometer is determined by the balance of \emph{only two} forces: an external force, $F_\text{ext}$, applied to the movable layer of the bilayer graphene, which should be measured, and a force of interlayer interaction, $F_\text{int}$. The feedback sensing of the external force is based on the dependence of the tunneling conductance $G$ on the displacement of the movable layer under the action of the external force.

\begin{figure}[!t]
\centering
\includegraphics{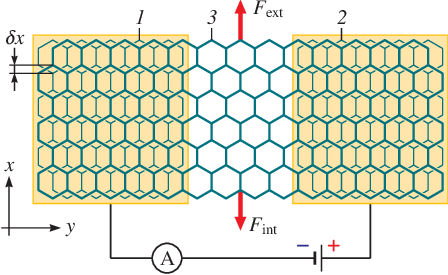}
\caption{Conceptual design of the nanodynamometer. The bottom graphene layers fixed on the electrodes are indicated as \emph{1} and \emph{2} and the movable
top graphene layer is indicated as \emph{3}.}
\end{figure}

The paper is organized in the following way. Calculations of the dependence of the force of interlayer interaction on the relative displacement of graphene layers and estimations of accuracy of force measurements are presented in Section II. Section III is devoted to calculations of the tunneling conductance between graphene layers. Our conclusions and discussion of possible applications of the nanodynamometer are summarized in Section IV.

\section{Interlayer interaction of bilayer graphene}
To study the dependence of the force of interlayer interaction, $F_\text{int}$, on the relative displacement of graphene layers the interlayer interaction of bilayer graphene has been investigated in the framework of the density functional theory with the dispersion correction (DFT-D).\cite{Grimme06, WuVargas01} The periodic boundary conditions are applied to a 4.26 \AA{} $\times$ 2.46 \AA{} $\times$ 20 \AA{} model cell. The VASP code \cite{Kresse96} with the density functional of Perdew, Burke, and Ernzerhof\cite{Perdew96} corrected with the dispersion term (PBE-D) \cite{Barone09} is used. The basis set consists of plane waves with the maximum kinetic energy of 800 eV. The interaction of valence electrons with atomic cores is described using the projector augmented-wave method (PAW).\cite{Kresse99} Integration over the Brillouin zone is performed using the Monkhorst--Pack method \cite{Monkhorst76} with $24\times36\times1$ $k$-point sampling. In the calculations of the potential energy reliefs, one of the graphene layers is rigidly shifted parallel to the other. Account of structure deformation induced by the interlayer interaction was shown to be inessential for the shape of the potential relief for the interaction between graphene-like layers, such as the interwall interaction of carbon nanotubes \cite{Kolmogorov00, Belikov04} and the intershell interaction of carbon nanoparticles.\cite{LozovikPopov00, LozovikPopov02} The DFT-D calculations show that the ground state of bilayer graphene corresponds to the AB stacking (Bernal structure) with the interlayer spacing $\delta Z = 3.25$ \AA{} and the interlayer interaction energy $-50.6$~meV$/$atom. 
The interaction of a single carbon atom in the graphene flake with the graphite surface was described using the simple approximation~\cite{Verhoeven04,Kerssemakers97} containing only the first Fourier components. Based on that expression, the interlayer interaction energy $U(\delta x, \delta y)$ as a function of the relative displacements $\delta x$ and $\delta y$ of the layers along the axes $x$ and $y$ chosen along the armchair and zigzag directions, respectively, at the equilibrium interlayer spacing can be roughly approximated in the form~\cite{Lebedeva10}
\begin{align}\label{eq101}
   U &= U_1 \biggl(1.5 + \cos\biggl(2k_x\delta x - \frac{2\pi}{3}\biggr) -{} \notag\\
   &- 2\cos\biggl(k_x\delta x - \frac{\pi}{3}\biggr)\cos(k_y\delta y)\biggr) + U_0,
\end{align}
where $k_y=2\pi/(\sqrt{3}a_\text{CC})$, $k_x=k_y/\sqrt{3}$, $a_\text{CC} = 1.42$ \AA{} is the bond length of graphene (see Fig.~2), $\delta x = 0$ and $\delta y = 0$ at the AB stacking. The parameters $U_0 = -101.18$ meV and $U_1 = 8.48$ meV (per elementary unit cell) are fitted to reproduce the potential energy relief of bilayer graphene. The relative root-mean-square deviation  $\delta U/U_1$ of approximation (\ref{eq101}) from the potential energy relief obtained using the DFT-D calculations is found to be $\delta U/U_1=0.043$. The potential energy relief calculated using approximation (\ref{eq101}) is shown in Fig.~3.

\begin{figure}[!t]
\centering
\includegraphics{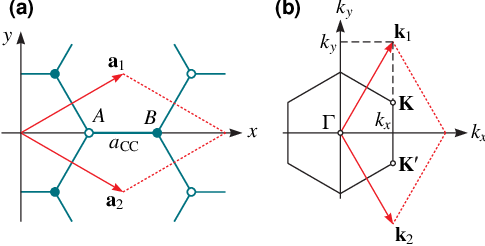}
\caption{Structure of a single layer of graphene (see, e.g., Ref.~\onlinecite{CastroNeto09}) in real space (a) and in reciprocal space (b). The elementary unit cell is denoted by dotted lines. The hexagon in part (b) is the boundary of the first Brillouin zone. $\vec{a}_1$ and $\vec{a}_2$ are the translational vectors, $\vec{k}_1$ and $\vec{k}_2$ are vectors reciprocal to $\vec{a}_1$ and $\vec{a}_2$; $k_x$ and $k_y$ are projections of $\vec{k}_1$ and $\vec{k}_2$ on coordinate axes. Nonequivalent lattice sites are denoted by $A$ and $B$.}
\end{figure}
\begin{figure}[!b]
\centering
\includegraphics{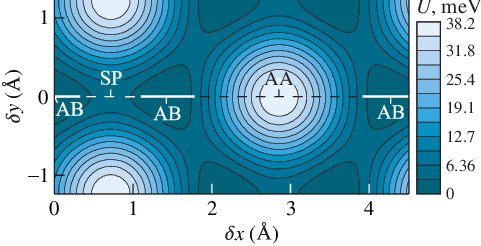}
\caption{Calculated interlayer interaction energy $U$ of bilayer graphene per elementary unit cell as a function of the relative position $\delta x$ and $\delta y$ of the layers. The energy is given relative to the global energy minimum $U_0$. SP is a saddle point in the potential relief. The regions where the stable equilibrium is possible and not possible on the displacement of the movable layer along the armchair direction are shown with the solid and dashed lines, respectively.}
\end{figure}

For simplicity, we restrict the analysis of operation of the nanodynamometer by the case where the external force $F_\text{ext}$ is directed along
the $x$ (armchair) direction, i.e. along the path between adjacent energy minima. The dependence of the interlayer force on the displacement of the
movable layer in this direction can be calculated using approximation (\ref{eq101}),
\begin{align}\label{eq102}
   F_\text{int} = -\frac{\partial U}{\partial x} = 2 U_1 k_x \biggl(\sin\biggl(2k_x\delta x - \frac{2\pi}{3}\biggr) -{} \notag\\
   -\sin\biggl(k_x\delta x- \frac{\pi}{3}\biggr)\cos(k_y\delta y)\biggr).
\end{align}

This dependence is shown in Fig.~4a. Under the action of the external force $F_\text{ext}$, the equilibrium position of the layers is determined by the condition $F_\text{ext} + F_\text{int} = 0$. This equilibrium is stable if the matrix of the second derivatives of the potential function $U(x,y)$ is positive definite. 
Differentiating Eq.~(\ref{eq101}), we find that the stable equilibrium is possible up to the displacement $|\delta x_1| \approx 0.229a_\text{CC}$ in the direction corresponding to transition from the AB stacking to the SP stacking and up to the displacement $|\delta x_2| = a_\text{CC}/4$ in the direction corresponding to transition from the AB stacking to the AA stacking. The maximum forces that can be measured in these directions are $F_1 = 15$~pN and $F_2 = 40$~pN per elementary unit cell, respectively. In Figs. 3 and 4, the regions where the stable equilibrium is possible and is not possible are shown with the solid the dashed lines, respectively.

\begin{figure}[!b]
\centering
\includegraphics{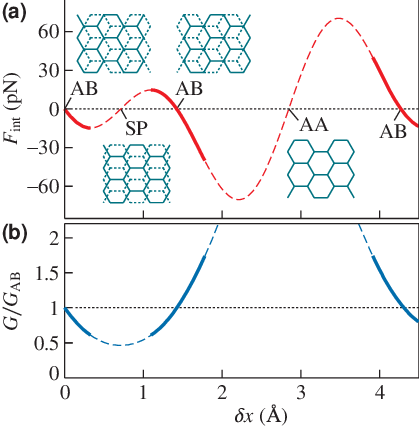}
\caption{(a) Calculated force $F_\text{int}$ of the interlayer interaction acting on the movable layer of bilayer graphene per elementary unit cell as a function of the relative displacement $\delta x$ of the movable layer in the $x$ (armchair) direction obtained using approximation (\ref{eq101}). The regions where the stable equilibrium is possible and not possible are shown with the solid and dashed lines, respectively. (b) Calculated tunneling conductance $G/G_\text{AB}$ between the layers as a function of the relative displacement $\delta x$ of the movable layer in the $x$ direction.}

\vspace{-4pt}
\end{figure}

The upper limit of forces that can be measured using the graphene-based nanodynamometer is proportional to the overlap area of the graphene layers
and is given by $F_\text{max} \approx F_2 N_\text{G}$, where $N_\text{G}$ is the number of the elementary unit cells in the overlap area. For example,
for the overlap areas of $10^2$ and $10^4$~nm$^2$, the maximum forces that can be measured are 76~nN and 7.6~$\mu$N, respectively. The
accuracy of the force measurements is limited by thermal vibrations of the graphene layers. The amplitude of these vibrations can be estimated as
\begin{equation}\label{eq103}
\langle x^2 \rangle_T \approx \frac{k_\text{B} T}{N_\text{G}} \left(\frac{\partial^2 U}{\partial x^2} \right)^{-1},
\end{equation}
where $k_\text{B}$ is the Boltzmann constant, $T$ is temperature and $\partial^2 U /\partial x^2$ is the second derivative of the interlayer interaction energy with respect to the displacement of the layers along the armchair direction at the energy minimum. The latter quantity is found to be equal $\partial^2U/\partial x^2 = 3U_1k_x^2 = 55.3$~meV$/$\AA$^2$ per elementary unit cell from Eq.~(\ref{eq101}). 
The relative error of the force measurements can be estimated as the ratio of the amplitude of the thermal vibrations to the maximal displacement of graphene layers where the stable equilibrium is possible. At liquid helium and room temperatures, these ratios equal $\sqrt{\langle x^2\rangle_T}/\delta x_2 = 0.23/\sqrt{N_\text{G}}$ and $\sqrt{\langle x^2\rangle_T}/\delta x_2 = 1.9/\sqrt{N_\text{G}}$, respectively. So for the overlap area of 100~nm$^2$, these quantities are 0.005 and 0.044, respectively. It is seen that the relative error of the force measurements decreases with increasing the overlap area of the graphene layers.

\section{Conductance of bilayer graphene}

Let us show that tunneling conductance $G$ between graphene layers changes considerably with the relative displacement of the layers and therefore measurements of the conductance $G$ can be used to determine this displacement. We use the Bardeen method,\cite{Bardeen61} which was previously applied for calculation of the tunneling conductance between walls of double-walled carbon nanotubes.\cite{PoklonskiHieu08} It is known \cite{Tersoff85} that the tunneling conductance is proportional to the sum of squares of the amplitudes of the tunneling transition (tunneling matrix elements) for all electron states at both sides of the tunneling transition.
This approach was used previously to study the electronic structure \cite{Bistritzer11} and conductance \cite{Bistritzer10} of twisted two-layer graphene system. Here we use such an approach to calculate the relative changes of tunneling conductance $G$ between the layers at their relative displacement from the ground state corresponding to AB stacking.

In the framework of the Bardeen's formalism,\cite{Bardeen61} the amplitude of the tunneling transition between states of the bottom ($\Psi_\text{bot}$) and top ($\Psi_\text{top}$) layers of bilayer graphene is given by
\begin{equation}\label{eq104}
   M_\text{bot,top}^{\vec{k}_\text{bot},\vec{k}_\text{top}} = \frac{\hbar^2}{2m_0}\int_S (\Psi_\text{bot}^{*} \nabla \Psi_\text{top} - \Psi_\text{top} \nabla \Psi_\text{bot}^{*})\,d\vec{S},
\end{equation}
where $S$ is the overlap area between the graphene layers, $\vec{k}_\text{bot}$ and $\vec{k}_\text{top}$ are two-dimensional vectors in the
reciprocal space of the graphene lattice corresponding to the bottom and top layers, $m_0$ is the electron mass in vacuum, $\hbar = h/2\pi$ is the Planck constant.

In the tight-binding approximation for vectors $\vec{k}_\text{bot}$ (or $\vec{k}_\text{top}$) near the corners ($K$-points,\cite{CastroNeto09}
$\vec{K} = (2\pi/(3a_\text{CC}), 2\pi/(3\sqrt{3}a_\text{CC}))$ and
$\vec{K}' = (2\pi/(3a_\text{CC}), -2\pi/(3\sqrt{3}a_\text{CC}))$) of the Brillouin zone (Fig.~2b), the wave function of the bottom graphene layer takes the form \cite{Barnett05}
\begin{align}\label{eq105}
   \Psi_\text{bot} &= \frac{1}{\sqrt{N_\text{G}}}\sum_{g=1}^{N_\text{G}}\exp(i\vec{k}_\text{bot}\vec{R}_g^\text{bot})\times{} \notag\\
   &\times\frac{1}{\sqrt{2}}\bigl(\chi(\vec{r} - \vec{R}_g^\text{bot}) \pm \chi(\vec{r} - \vec{R}_g^\text{bot} - \vec{d})\bigr),
\end{align}
and the same formula for $\Psi_\text{top}$.
Here $N_\text{G}$ is the number of the elementary unit cells of graphene, $\vec{d}$ is the vector between two non-equivalent carbon atoms ($A$ and $B$) in the elementary unit cell, $d = a_\text{CC}$; signs $+$ and $-$ correspond to $\pi$- (bonding) and $\pi^*$- (antibonding) orbitals in graphene, respectively, $\vec{R}_g^\text{bot}$ is the radius vector of the $g$-th unit cell of the bottom graphene layer, $\vec{r}$ is the radius vector. The function $\chi(\vec{r})$ is a Slater $2p_x$-orbital
\begin{equation}\label{eq106}
   \chi\left(\vec{r}\right) = \left(\frac{\xi^5}{\pi}\right)^{1/2} z\,\exp\left(-\xi\sqrt{x^2 + y^2 + z^2}\right),
\end{equation}
where \cite{Clementi63} $\xi=1.5679/a_\text{B}$ and $z$ is the axis perpendicular to the graphene plane; $a_\text{B} = 0.529$ \AA{} is the Bohr radius and $r = \sqrt{x^2 + y^2 + z^2}$ is the magnitude of the radius vector $\vec{r}$ from carbon atom center.

Let us substitute the wave function (\ref{eq105}) into Eq.~(\ref{eq104}).
The product $\Psi_\text{bot}^*\nabla\Psi_\text{top}$ in Eq.~(\ref{eq105}) can be rewritten as
\begin{align}\label{eq116}
   &\Psi_\text{bot}^*\nabla\Psi_\text{top} = \notag\\
   &= \frac{1}{2N_\text{G}} \sum_{g=1}^{N_\text{G}}\exp(-i\vec{k}_\text{bot}\vec{R}_g^\text{bot})\bigl(\chi(\vec{r} - \vec{R}_g^\text{bot}) \pm \chi(\vec{r} - \vec{R}_g^\text{bot} - \vec{d})\bigr)\times{} \notag\\
   &\times \nabla\sum_{g'=1}^{N_\text{G}}\exp(i\vec{k}_\text{top}\vec{R}_{g'}^\text{top}) \bigl(\chi(\vec{r} - \vec{R}_{g'}^\text{top}) \pm \chi(\vec{r} - \vec{R}_{g'}^\text{top} - \vec{d})\bigr)= \notag\\
   &= \frac{1}{2N_\text{G}} \sum_{g=1}^{N_\text{G}}\sum_{h=1}^{N_\text{G}} \exp(i(\vec{k}_\text{top}\vec{R}_g^\text{bot} - \vec{k}_\text{bot}\vec{R}_g^\text{bot}) + i\vec{k}_\text{top}\Delta\vec{R}_h)\times{} \notag\\
   &\times \bigl(\chi(\vec{r}') \pm \chi(\vec{r}' - \vec{d})\bigr) \nabla\bigl(\chi(\vec{r}' - \Delta\vec{R}_h) \pm \chi(\vec{r}' - \Delta\vec{R}_h - \vec{d})\bigr),
\end{align}
where $\vec{r}' = \vec{r} - \vec{R}_g^\text{bot}$.
In Eq.~(\ref{eq116}) the coordinates of unit cells of the top layer $\vec{R}_{g'}^\text{top} = \vec{R}_g^\text{bot} + \Delta \vec{R}_h$ are expressed via the coordinates of unit cells of the bottom layer $\vec{R}_g^\text{bot}$ and displacements of unit cells of the top layer $\Delta \vec{R}_h$.

Since all unit cells of graphene are identical, only one cell of the bottom layer can be considered in the calculation of $M_\text{bot,top}^{\vec{k}_\text{bot},\vec{k}_\text{top}}$. It should also be taken into account that for large $N_\text{G} \gg 1$ the following relation is satisfied
\begin{equation}\label{eq107}
   \frac{1}{N_\text{G}} \sum_{g=1}^{N_\text{G}} \exp(i \vec{k}_\text{top} \vec{R}_g)
\exp(-i \vec{k}_\text{bot} \vec{R}_g)
= \delta_{\vec{k}_\text{bot},\vec{k}_\text{top}},
\end{equation}
where $\delta_{\vec{k}_\text{bot},\vec{k}_\text{top}}$ is the Kronecker symbol. 

Taking into account Eqs.~(\ref{eq116}) and (\ref{eq107}), the Eq.~(\ref{eq104}) takes the form
\begin{align}\label{eq108}
   &M_\text{bot,top}^{\vec{k}_\text{bot},\vec{k}_\text{top}} = \frac{\hbar^2}{2m_0} \sum_{g=1}^{N_\text{G}} \frac{1}{2N_\text{G}}\exp(i \vec{k}_\text{top}\vec{R}_g^\text{bot} - i \vec{k}_\text{bot}\vec{R}_g^\text{bot}) \times{} \notag\\
   &\times \sum_{h=1}^{N_\text{G}}\exp(i \vec{k}_\text{top}\Delta\vec{R}_h) \int_S \bigl(\chi(\vec{r} - \Delta\vec{R}_h) \pm \chi(\vec{r} - \Delta\vec{R}_h - \vec{d})\bigr) \times{} \notag \\
   &\times \nabla\bigl(\chi(\vec{r}) \pm \chi(\vec{r} - \vec{d})\bigr)\,d\vec{S} = M_\text{bot,top}^{\vec{k}_\text{top}} \delta_{\vec{k}_\text{bot},\vec{k}_\text{top}}.
\end{align}
Applying $\delta_{\vec{k}_\text{bot},\vec{k}_\text{top}}$ in Eq.~(\ref{eq108}), we get only vectors $\vec{k}_\text{top} = \vec{k}_\text{bot}$. This yields for $\vec{k}_\text{top} = \vec{k}_\text{bot} = \vec{K}$ (or $\vec{k}_\text{top} = \vec{k}_\text{bot} = \vec{K}'$; see Fig.~2b):
\begin{align}\label{eq109}
   &M_\text{bot,top}^{\vec{k}_\text{bot},\vec{k}_\text{top}} = M_\text{bot,top}^{\vec{K},\vec{K}} = M_\text{bot,top}^{\vec{K}',\vec{K}'} = \frac{\hbar^2}{2m_0} \sum_{h=1}^{N_\text{G}} \frac{1}{2}\exp(i \vec{K}\Delta\vec{R}_h) \times{} \notag\\ 
   &\times \int_S \bigl(\chi(\vec{r} - \Delta\vec{R}_h) \pm \chi(\vec{r} - \Delta\vec{R}_h - \vec{d})\bigr) \nabla\bigl(\chi(\vec{r}) \pm \chi(\vec{r} - \vec{d})\bigr)\,d\vec{S} \approx{} \notag\\
   &\approx \sum_{h=1}^{n_\text{G}}\exp(i \vec{K}\Delta\vec{R}_h) (\gamma_{A-A'_h}+\gamma_{A-B'_h}+\gamma_{B-A'_h}+\gamma_{B-B'_h}),
\end{align}
where $n_\text{G} = [\pi\Delta R_\text{max}^2/(\sqrt{3}a_1^2/2)]$ is the number of unit cells of the top layer located at the distance in graphene plane less than $\Delta R_\text{max}$ from the considered unit cell of the bottom layer, $\gamma_{A(B)-A'_h(B'_h)}$ are the hopping integrals between atom $A$ (or $B$) in the considered unit cell of the bottom layer and atom $A'_h$ (or $B'_h$) in the $h$-th unit cell of the top layer.
We use $\Delta R_\text{max} = 2a_1 = 2a_2 = 2\sqrt{3}a_\text{CC}$ and $n_\text{G} = [8\pi/\sqrt{3}] = 14$ in the calculations for the both layers (Fig.~2a), 
taking into account that the interactions between atoms lying at longer distances change the value of the matrix element by less than~0.1\%.

The hopping integrals $\gamma_{A(B)-A'_h(B'_h)}$ in Eq.~\eqref{eq109} are given by
\begin{align}\label{eq110}
   \hspace{-6pt}\gamma_{A(B)-A'_h(B'_h)} &= \gamma_\rho(x, y) \notag\\
   &= \frac{\hbar^2}{2m_0} \int_S \frac{1}{2}\left(\chi_\text{bot}\frac{d}{dz}\chi_\text{top} - \chi_\text{top}\frac{d}{dz}\chi_\text{bot}\right)dS,
\end{align}
where the index $A(B)$ denotes atom $A$ (or atom $B$), $\chi_\text{bot} = \chi(x - X_{A(B)}, y - Y_{A(B)}, -\delta Z/2)$, $\chi_\text{top} = \chi\bigl(x - (X_{A'_h(B'_h)} + \delta x),$ $y - (Y_{A'_h(B'_h)} + \delta y), \delta Z/2\bigr)$, $X_{A(B)}$ and $Y_{A(B)}$ are the coordinates of atom $A$ (or atom $B$) in the elementary unit cell of the bottom layer ($X_A = 0$, $X_B = a_\text{CC}$, $Y_A = Y_B = 0$), $X_{A'_h(B'_h)}$ and $Y_{A'_h(B'_h)}$ are the coordinates of atoms of the top layer for bilayer graphene in the ground state (AB stacking), and $\delta Z/2 = 1.625$~\AA{} is the half of the interlayer distance, 
$\bm{\rho} = (X_{A'_h(B'_h)} + \delta X - X_{A(B)}$, $Y_{A'_h(B'_h)} + \delta Y - Y_{A(B)}$, $0)$ is the projection on graphene plane of the vector connecting two selected atoms in bilayer graphene;
for $\gamma_{A-A'_h}$ and $\gamma_{B-B'_h}$ vector $\bm{\rho}$ is the projection of $\Delta \vec{R}_h$,
for $\gamma_{A-B'_h}$ vector $\bm{\rho}$ is the projection of $\Delta\vec{R}_h+\vec{d}$, and
for $\gamma_{B-A'_h}$ vector $\bm{\rho}$ is the projection of $\Delta\vec{R}_h-\vec{d}$.
Analogously to Refs.~\onlinecite{Bistritzer11, Bistritzer10} the hopping integral $\gamma_\rho$ depends on the magnitude $\rho$ of vector $\bm{\rho}$ (see Fig.~5). The function $\gamma_\rho$ can be approximated (with accuracy within 3\%) by an expression $\gamma_\rho = \gamma_\text{max}\exp(-\zeta(\rho/a_\text{CC})^2)$, where $\gamma_\text{max} = 189$~meV and $\zeta = 0.8$.

The calculation according to Eq.~(\ref{eq110}) for the AA stacking of bilayer graphene, in which equivalent atoms of the top and bottom layers are located opposite to each other, gives the values of the hopping integrals $\gamma_{A'-A} = \gamma_{B'-B} = 189.2$~meV.
The expressions (\ref{eq106}), (\ref{eq109}) and (\ref{eq110}) allow to calculate the tunneling matrix element.
For the AB stacking (the Bernal structure, $\delta x=0$ and  $\delta y=0$), the amplitude of the tunneling transition between the states $\Psi_\text{bot}$ and $\Psi_\text{top}$ of the bottom and top layers of bilayer graphene is found to be $|M_\text{bot,top}^{\vec{K},\vec{K}}| \approx 136$~meV, while for the AA stacking ($\delta x = -a_\text{CC}, \delta y = 0$), $|M_\text{bot,top}^{\vec{K},\vec{K}}| \approx 272$~meV (see Eq.~(\ref{eq109})).

\begin{figure}[!t]
\centering
\includegraphics{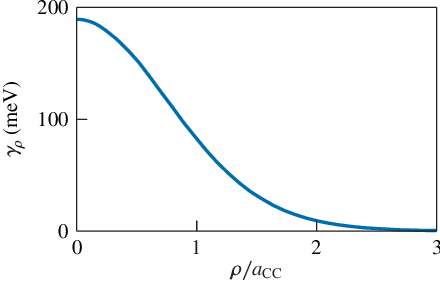}
\caption{The hopping integral $\gamma_\rho$ as a function of the magnitude $\rho$ of vector $\bm{\rho}$ (see Eq.~(\ref{eq110}))}
\end{figure}

The ratio of the tunneling conductance $G$ to the tunneling conductance $G_\text{AB}$ of bilayer graphene at the ground state ($\delta x = 0$) equals to the ratio $|M_\text{bot,top}^{\vec{K},\vec{K}}|^2/|M_\text{bot,top}^{\vec{K},\vec{K}}|_{\delta x=0}^2$ determined by Eq.~(\ref{eq109}). The dependence of this ratio on the relative displacement of the layers along the $x$ (armchair) direction is shown in Fig.~4b. It is seen that the tunneling conductance between the graphene layers strongly depends on their relative position at the sub-nanometer scale, similar to the results obtained for double-walled carbon nanotubes.\cite{Grace04, Tunney06, PoklonskiHieu08} The conductance reaches its maximum for the AA stacking, in which atoms of the layers are located at the smallest distances to each other. The minimum of the tunneling conductance corresponds to the SP stacking. Figure 4 shows that the relative displacement of the graphene layers in the course of the operation of the nanodynamometer can result in changes of the tunneling conductance $G$ in the relatively wide range from 0.61$G_\text{AB}$ to 1.73$G_\text{AB}$. Thus it is seen that the relative displacement $\delta x$ of the layers (Fig.~4b) and, consequently, the external force acting to the layers (Fig.~4a) can be determined by the measurements of the electrical conductance between the layers.

The model that we use to calculate the tunneling conductance adequately describes electron tunneling for relative positions of the graphene layers in which their atoms are not located exactly opposite to each other. In the case when the atoms are located exactly opposite to each other, hybridization of their wave functions occurs leading to a significant increase of the conductance which is now determined not by tunneling between the layers but rather by transitions between energy bands of the combined electron system of bilayer graphene. Therefore our calculations provide only the lower bound estimate of the relative variation of the tunneling conductance upon the relative displacement of the graphene layers. Nevertheless even this estimate is sufficient to demonstrate the feasibility of force measurements using the proposed design of the nanodynamometer.

Let us also consider the possibility of a nanodynamometer based on the relative rotation of graphene layers. At a relative translational displacement of the layers from the ground state corresponding to the AB stacking, the interlayer interaction energy increases and the tunneling conductance between the layers increases or decreases (depending on direction of displacement) identically for all local areas of the overlap. Contrary to that case, at a relative rotation of the layers, the interlayer interaction energy and the tunneling conductance change differently for local areas of the overlap. While the interlayer interaction energy increases for any local area since the AB stacking corresponds to the global energy minimum, the tunneling conductance increases for some local areas and decreases for the others. As a result, contributions from different local areas to the total tunneling conductance compensate each other. Therefore changes in the total tunneling conductance at the relative rotation of the layers are much smaller than such changes at the relative translational displacement. Moreover the force required for the relative rotation of the layers from the AB stacking to the incommensurate state is an order of magnitude greater than the force required for the displacement from the AB stacking to the SP stacking.\cite{Lebedeva10}
Thus the scheme of the nanodynamometer based on the relative rotation of the graphene layers is less effective than the proposed scheme based on the relative displacement of the layers. For the proposed scheme of the nanodynamometer, the relative rotation of the layers should be avoided, i.e. only forces that do not produce a significant torque should be considered. This is the case when the measured force acts 1) uniformly on all atoms of the upper layer, 2) on adsorbents uniformly distributed on the surface or edges of the upper layer in the area between the first and second bottom layers 3) on a nanoobject placed near the center of the upper layer.

\section{Discussion and Conclusions}

We have proposed the concept of the electromechanical nanodynamometer based on bilayer graphene in which the force is determined by measurements of the conductance between the layers. In this nanodynamometer, the force acting on one of the graphene layers causes the relative displacement of this layer and related change of the conductance between the layers. The calculations of the potential relief of the interlayer interaction energy within the dispersion-corrected density functional theory approach showed that the stable equilibrium of bilayer graphene is possible if the measured force acting on one of the layers along the armchair direction does not exceed 40~pN per elementary unit cell. The corresponding displacement of graphene layers lies within 0.36~\AA{}. 
The calculations of the tunneling conductance of bilayer graphene using the Bardeen method allowed us to estimate that on the relative displacement of the layers, the tunneling conductance changes by at least a factor of 2, which provides the excellent possibility to determine the force by the conductance measurements.
The relative error of the force measurements is determined by the relative thermal vibrations of the layers. This error decreases with the increase of the overlap of the layers and with the decrease of temperature.

Let us discuss possible applications of the considered nanodynamometer. A molecule or a nanoobject can be adsorbed on the top layer of the
nanodynamometer in the region where the top layer does not overlap with the bottom layers. The measurements of the force acting on the molecule or
nanoobject in the presence of an electric or magnetic field would allow to determine their polarizability, electric and magnetic dipole and
quadrupole moments.

In the pioneering work of Novoselov \emph{et al} graphene flakes were placed on an insulating substrate and brought into contact with electrodes \cite{Novoselov04} (to create field-effect transistor). A further considerable progress has been achieved in manipulation of individual graphene layers. Individual graphene flakes were moved on a graphite surface by the tip of the friction force microscope.\cite{Dienwiebel04} The possibilities to cut graphene nanoribbons with desirable geometrical parameters \cite{LiuZhang11} and remove individual graphene layers in a controllable way for device patterning \cite{Dimiev11} were demonstrated. The tunneling conductance can be measured for graphene in the way similar to the experiments for multiwall carbon nanotubes.\cite{Stetter10, Bourlon04} All these give us a cause for optimism that the proposed graphene-based electromechanical nanodynamometer will be implemented in the near future.

%% Acknowledgement section
\begin{acknowledgments}
This work has been partially supported by the RFBR (grant 12-02-90041-Bel, 11-02-00604-a) and BFBR (grant Nos. F11V-001, F12R-178). The DFT-D calculations have been performed on the SKIF MSU Chebyshev supercomputer and on the MVS-100K supercomputer at the Joint Supercomputer Center of the Russian Academy of Sciences.
\end{acknowledgments}

% Create the reference section using BibTeX:
%\bibliographystyle{vak}
\bibliography{GrapheneBasedNanodynamometer}

\end{document}